\documentclass[aps,prd,preprint]{revtex4-1}
\usepackage{tikz}
\usepackage {graphicx}
\usepackage[toc,page]{appendix}
\newcommand{\be}{\begin{equation}}
\newcommand{\ee}{\end{equation}}
\newcommand{\bea}{\begin{eqnarray}}
\newcommand{\eea}{\end{eqnarray}}
\begin{document}
\title{Relativity for Retired Engineers}
\author{David Garfinkle}
\email{garfinkl@oakland.edu}
\affiliation{Dept. of Physics, Oakland University, Rochester, MI 48309, USA}
\affiliation{Leinweber Center for Theoretical Physics, Randall Laboratory of Physics, University of Michigan, Ann Arbor, MI 48109-1120, USA}


\date{\today}

\begin{abstract}
We provide some guidance and examples to clear up common misconceptions about special relativity.  These misconceptions often come from trying to express the truths of special relativity in Newtonian terms rather than in terms more natural to special relativity itself.
This conceptual stance can also help in attaining a better understanding of general relativity.

\end{abstract}


\maketitle

\section{Introduction}
Scarcely a month goes by when I don't receive an email, phone call, or manuscript from someone, more often than not a retired engineer, who has their own theory of relativity.  These people have read something about Einstein's theory of relativity, noticed that something about it doesn't accord with their own intuition, decided that therefore Einstein must be wrong, and come up with their own theory which they are convinced must be right.  It is certainly true that there are aspects of relativity that are counterintuitive, not only to retired engineers but also to many physicists.  Therefore if one's own intuition were to be the final arbiter of correctness of a theory, that would be good reason to reject relativity.  But of course science doesn't work that way: the final arbiter for correctness of a theory is experimental evidence.  So if a counterintuitive theory is supported by the evidence, one should retrain ones intuition to think in the ways required by the theory. 

It is beyond the scope of even a moderate sized article to clear up the myriad misconceptions there are about general relativity; but it is possible, and is indeed the aim of this article, to clear up misconceptions about special relativity.  This will be salutary not only to those retired engineers convinced that special relativity must be wrong, but also to many physicists who while nominally accepting relativity are nonetheless convinced in their heart of hearts that their own laboratory is a preferred frame of rest.   

Section \ref{Babylon} explains why engineers are especially prone to thinking that there is something wrong with relativity, while sections \ref{boosts} and \ref{electrodynamics} present special relativity on its own terms.  Section \ref{clocks1} presents time dilation and length contraction, while section \ref{paradox} explains the classic paradoxes of special relativity.  Finally, section \ref{GR} explains how the stance of this paper can be used as a starting point for understanding general relativity.

\section{Use ``Babylonian mathematics'' with care}
\label{Babylon}
The typical retired engineer learned Euclidean geometry in high school, and Newtonian mechanics in college.  A long and productive career ensued, filled with practical measurements and calculations.  This practical career left no time for anything as esoteric as Einstein's theory of relativity. But now in retirement there is time, and surely the vast physical intuition built up in a lifetime of practical measurement and calculation is just what is needed to understand any physical theory.  What could go wrong?

Quite a lot actually.  In {\it The Character of Physical Law}\cite{feynman}, Richard Feynman made a distinction between what he called ``Greek mathematics'' and what he called ``Babylonian mathematics.''  Feynman was many things, but ``serious scholar of history'' was not one of them, and these terms of his are meant to be suggestive rather than historically accurate.  But the point that he used them to make is serious and important: Feynman noted that in Greek mathematics, a certain small set of truths are elevated to the highest position and called axioms.  All other truths are then to be derived from the axioms.  In contrast in Babylonian mathematics all truths are on an equal footing, and whichever set of truths one remembers, provided it is large enough, can be used to derive the rest.  Feynman claimed that while contemporary mathematicians may do Greek mathematics, physicists, especially those working out a new area of theory, use Babylonian mathematics.  

However, there is a pitfall to doing Babylonian mathematics: among the set of ``truths'' that one thinks one knows, there may be some that aren't actually true.  And then deriving things from these ``truths'' can lead one into ever more error.  Thus Babylonian mathematics must be used with caution: or as the old saying (usually misatributed to Mark Twain) goes, ``It aint what you don't know that gets you into trouble.  It's what you know for sure that just aint so.''

In particular, since Newtonian mechanics is inconsistent with special relativity, the retired engineer starting with the ``truths'' of Newtonian mechanics will eventually reason his way to something that contradicts special relativity, and will then conclude that special relativity is wrong.  Nor does it help to point out to them that Einstein's postulate that the speed of light is the same for all observers\cite{einstein1905} is already clearly in conflict with Newtonian mechanics: in response the retired engineer will simply introduce some ad hoc postulate of his own (since as far as he can see, ``that's how the game is played'') and derive even more contradictions with special relativity.

And where do physicists learn the appropriate caution to use with Babylonian mathematics?  {\emph {By learning special relativity}}.  We first learn Newtonian mechanics and then only later learn special relativity.  But learning special relativity requires unlearning some of Newtonian mechanics. In particular, we learn that the ``truths'' of Newtonian mechanics are only approximations that hold for speeds small compared to the speed of light.  

Thus, retired engineers are singularly ill suited to learning and understanding special relativity because of their tendency to do some or all of the following: (1) Charge ahead with Babylonian mathematics without ever having learned its pitfalls; (2) Trust the things they shouldn't trust (their own physical intuition); (3) Distrust the things they should trust (the things that physicists say about the well tested and therefore well established theory of special relativity);  (4) Think that a well established and well tested theory is in need of reformulation.

And yet, it should also be acknowledged that the standard textbook ways of presenting special relativity do make the subject seem paradoxical in the following way: on the one hand the effects being presented seem to be effects of speed.  Time dilation is ``fast clocks run slow'' while length contraction is ``fast rulers get short.''  On the other hand we are told that there is no fact of the matter about who is moving and if I think you are moving then you think I am moving.  Combining the two sets of assertions we then obtain the discomforting thought that ``you think my clocks run slow, but I think your clocks run slow'' and ``you think my rulers are short but I think your rulers are short.''  This discomfort is then made acute by paradoxes such as the ``twin paradox'' and the ``pole in the barn paradox'' which are then finally resolved.  At times one feels as though one is learning Zen Buddhism rather than physics.

Thus there may be room for improvement in the way that special relativity is often presented, not only for the sake of retired engineers but also for the sake of physics students.  How then should one present this subject?
Oddly enough it seems to me that the best method is to get them out of their comfort zone right away in the following sense: (1) Introduce as soon as possible the notions of spacetime, spacetime diagrams, the spacetime interval, four-vectors, proper time, and invariant quantities.  (2) Insist that for clarity all questions be formulated in terms of invariant quantities, and when needed accompanied by a spacetime diagram.  (3) Present special relativity on its own terms and do some physics with it first.  (4) Then (and only then) present the ``paradoxes'' of special relativity, but with the attitude that they are not paradoxical at all and simply result from trying to do new physics entirely in the language and frame of mind of the old physics.

In this way the retired engineer will not be tempted to try to state everything in a Newtonian formulation and draw mistaken conclusions from Newtonian ``truths.''  The rest of this article is an attempt to present special relativity according to the plan outlined above.  

I should mention that there are several treatments of special relativity 
(see {\it e.g.} \cite{wald1,gerochab,tevian}) that collectively do much of what I am calling for.  Thus, one can think of this manuscript as collecting various pieces of good special relativity pedagogy in one place.  But in addition, having written this manuscript, now the next time that I am contacted by a retired engineer with his own theory of relativity I can say something to the effect of ``I will gladly read your paper provided that at the same time you read this one and then we can discuss both.''  Thus I offer this manuscript not only to retired engineers but also to my fellow general relativists so that when {\emph {they}} are contacted by retired engineers they can respond in the same way.

\section{Boosts are like rotations}
\label{boosts}
Einstein did not invent the principle of relativity. That was done by Galileo.  Essentially the principle of relativity says that you can't feel speed, only changes in velocity.  Thus for two observers each at constant velocity there is no fact of the matter about who is moving and who is at rest. And indeed certain basic facts about astronomy are enough to convince us that something like that must be true: due to the Earth's rotation a person at the equator is moving at about 460 m/s.   Due to the Earth's orbit around the Sun, we are moving at about 30,000 m/s.  Not to mention the motion of the Sun within the galaxy and the motion of the galaxy through intergalactic space.  All this rapid motion and we don't feel a thing!  Indeed it was precisely because Copernican astronomy said that the apparent motion of the Sun and stars was due to Earth's rotation around its axis and its revolution around the Sun that Galileo felt the need to introduce the principle of relativity.

The principle of relativity doesn't mean that the point of view of all observers is equally valid: rather there is a special class of observers, the inertial observers, who travel at constant velocity.  But the point of view of all inertial observers {\emph {is}} equally valid: among the inertial observers there is no particular one who is really at rest with all the others really moving.  Because of this equality among inertial observers, there is no fact of the matter about what the position of an object is: an object that is seen to be at position $\vec x$ at time $t$ by one inertial observer will be seen to be at a different position by another. Thus we need some formula that allows us to go from the position $\vec x$ as seen by one observer to the position ${\vec x}'$ as seen by an observer in motion with velocity $\vec v$ relative to the first observer.  

For simplicity we start with one dimensional motion along a line, so that our positions and velocities are numbers rather than vectors.  In Galilean relativity the formula that relates $x'$ to $x$ is
\be
{x'}=x-vt \; \; \; .
\label{galileo}
\ee 

In Einstein's special relativity, different observers ascribe not only different positions but also different times.  In special relativity eqn. (\ref{galileo}) is replaced with
\bea
{x'}&=& \gamma (x-vt) \; \; \; ,
\\
{t'} &=& \gamma \left ( t - {\frac {vx} {c^2}} \right ) \; \; \; .
\label{lorentz1}
\eea
In section \ref{clocks1} we will give an operational reason for the transformation having this particular form, but for now we simply present it as a fact about physics.
This transformation is called a Lorentz transformation and is also sometimes callled a boost.
Here $c$ is the speed of light and $\gamma$ is given by
\be
\gamma ={\frac 1 {\sqrt {1-{\frac {v^2} {c^2}}}}}
\ee  
We can make these equations look more symmetric, and also give the transformed quantities the same units, by writing the formulas in terms of $x$ and $ct$ and their primed quantities.
\bea
{x'}&=& \gamma \left ( x-{\frac v c} ct \right ) \; \; \; ,
\\
{ct'} &=& \gamma \left ( ct - {\frac v c} x \right ) \; \; \; .
\label{lorentz2}
\eea
Even better, we can (and will) adopt units in which $c=1$, for example by measuring time in seconds and distance in light seconds. In that case the formula for $\gamma$ becomes
$\gamma = 1/{\sqrt {1-{v^2}}}$ while the formula for the boost becomes
\bea
{x'}&=& \gamma  ( x-v t  ) \; \; \; ,
\label{lorentz3a}
\\
{t'} &=& \gamma  ( t - v x ) \; \; \; .
\label{lorentz3b}
\eea

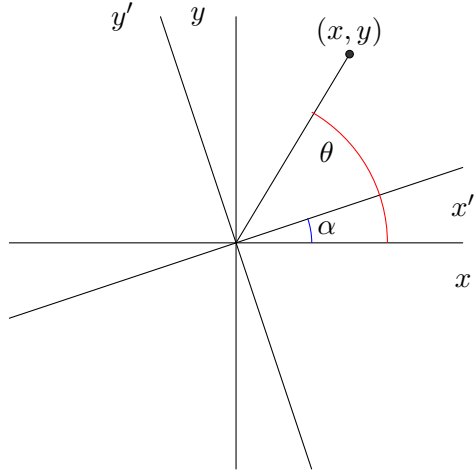
\begin{figure}
\centering
\begin{tikzpicture}
\draw (-3,0) -- (3,0);
\draw (0,-3) -- (0,3);
\draw (-3,-1) -- (3,1);
\draw (1,-3) -- (-1,3);
\draw (0,0) -- (1.5,2.5);
\draw[fill=black!80] (1.5,2.5) circle (0.05);
\node at (3,-0.5) {$x$};
\node at (-0.5,3) {$y$};
\node at (3,0.5) {${x'}$};
\node at (-1.5,3) {${y'}$};
\draw[blue] (1,0.0) arc
[
start angle=0,
end angle=19,
x radius = 1,
y radius =1
];
\draw[red] (2,0.0) arc
[
start angle=0,
end angle=60,
x radius = 2,
y radius =2
];
\node at (1.5,2.8) {$(x,y)$};
\node at (1.2,0.2) {$\alpha$};
\node at (1.2,1.2) {$\theta$};
\end{tikzpicture}
\caption{a point in the plane seen in $xy$ and ${x'}{y'}$ coordinate systems with the $x'$ axis at an angle $\alpha$ to the $x$ axis}
\label{fig1}
\end{figure}

The form of eqns. (\ref{lorentz3a}) and (\ref{lorentz3b}) looks similar to an elementary formula from analytic geometry: recall that in setting up the $x$ axis and $y$ axis in the plane, we can choose the $x$ axis to be in any direction, with the $y$ axis then perpendicular to that direction.  Now suppose that I choose a particular $x$ axis and you choose a different one at an angle $\alpha$ to mine. (see figure \ref{fig1}).  Then a point that I describe with the coordinates $(x,y)$ will be described by you with the coordinates $({x'},{y'})$.  What then is the formula for $({x'},{y'})$ in terms of $(x,y)$? Since $(x,y)=(r\cos \theta,r\sin \theta)$ and 
$({x'},{y'})=(r\cos (\theta - \alpha),r\sin (\theta-\alpha))$ we find

\bea
{x'}&=& \cos \alpha \, x + \sin \alpha \, y \; \; \; ,
\label{rotationa}
\\
{y'} &=& \cos \alpha \, y - \sin \alpha \, x \; \; \; .
\label{rotationb}
\eea

Note that ${{x'}^2}+{{y'}^2}={x^2}+{y^2}$, (as must be the case since each of these quantities is the square of the distance of the point from the origin) and that this identity comes about because of the trigonometric identity ${\cos^2}\alpha+{\sin^2}\alpha=1$.  The coefficients in the boost, $\gamma$ and $\gamma v$, don't satisfy this identity, but they satisfy a very similar one
\be
{\gamma ^2} - {{(\gamma v)}^2} = {\gamma ^2}(1-{v^2}) = 1 \; \; \; .
\ee 
Now recall the hyperbolic cosine ($\cosh$) and hyperbolic sine ($\sinh$) defined by 
$\cosh \psi \equiv ({e^\psi}+{e^{-\psi}})/2$ and $\sinh \psi \equiv ({e^\psi}-{e^{-\psi}})/2$.  These functions satisfy the identity
\be
{\cosh ^2} \psi - {\sinh ^2} \psi = {\textstyle {\frac 1 4}} ( {e^{2\psi}} + {e^{-2\psi}} + 2) - {\textstyle {\frac 1 4}} ( {e^{2\psi}} + {e^{-2\psi}} - 2) = 1 \; \; \; .
\ee
Thus there is some $\psi$ for which $\gamma = \cosh \psi $ and $\gamma v = \sinh \psi$.  With the usual definition of $\tanh$ that $\tanh \psi = \sinh \psi /\cosh \psi$ we then find that 
$\tanh \psi = v$ which in turn means that $\psi = {\tanh ^{-1}}v$.  The boost can then be written as
\bea
{x'}&=& \cosh \psi \, x-\sinh \psi \, t \; \; \; ,
\label{lorentz4a}
\\
{t'} &=& \cosh \psi \, t - \sinh \psi \, x \; \; \; .
\label{lorentz4b}
\eea 
Comparing eqns. (\ref{lorentz4a}) and (\ref{lorentz4b}) to eqns. (\ref{rotationa})
and (\ref{rotationb}) we see that boosts are like rotations.

Note that while eqns. (\ref{rotationa}) and (\ref{rotationb}) are a basic fact of the analytic geometry used in Newtonian physics, we don't tend to belabor the point.  That is, we don't spend a lot of time thinking about how my $x$ axis is different from your $x$ axis, and we don't spend a lot of time calculating transformations of various quantities from my $(x,y)$ coordinates to your $({x'},{y'})$ coordinates.  Thus one consequence of ``boosts are like rotations'' is the notion that traditional textbook emphasis on teaching special relativity by doing many calculations using boosts may be misplaced: not getting to the heart of the subject and making it look overly complicated.

For two points in the plane $({x_1},{y_1})$ and $({x_2},{y_2})$ we define 
$\Delta x \equiv {x_2}-{x_1}$ and $\Delta y \equiv {y_2}-{y_1}$.  The square of the distance
$\Delta \ell$ between the two points is 
\be
\Delta {\ell ^2} = \Delta {x^2} + \Delta {y^2} \; \; \; .
\label{length}
\ee
The formula for this distance is unchanged under a rotation.  

Similarly consider two ``points'' $({t_1},{x_1})$ and $({t_2},{x_2})$.  Though we will often refer to them as points, each such point represent a single point in space at a single moment in time.  Therefore such points are often called ``events'' and the two dimensional $tx$ space to which they belong is called ``spacetime.''  What ``squared distance'' is unchanged under a boost?  It is called the spacetime interval $\Delta {s^2} $ and is given by the formula
\be
\Delta {s^2} = -\Delta {t^2} + \Delta {x^2} \; \; \; .
\label{interval1}
\ee

Note that this notation is a little bit misleading because the quantity $\Delta {s^2}$ can be positive, negative, or zero.  What is the physical meaning of $\Delta {s^2}$ in each of these cases? When $\Delta {s^2} = 0$, this means that $|\Delta x/\Delta t|=1$ so a light ray could start at the event $({t_1},{x_1})$ and end at the event $({t_2},{x_2})$.  Thus the fact that all observers agree on the speed of light is simply a consequence of the fact that the spacetime interval is invariant.  (If all observers agree on the value of the spacetime interval then all observers agree on when it is zero).  If  $\Delta {s^2} < 0$, this means that $|\Delta x/\Delta t|<1$ so an object traveling at a constant velocity slower than light could start at the event $({t_1},{x_1})$ and end at the event $({t_2},{x_2})$.  In this case the time elapsed on a clock carried by such an object would be $\Delta \tau = {\sqrt {- \Delta {s^2}}}$.  The time elapsed on the clock is called proper time.  If  $\Delta {s^2} > 0$, this means that $|\Delta x/\Delta t|>1$ so no object or light ray could start at the event $({t_1},{x_1})$ and end at the event $({t_2},{x_2})$.  However, it turns out that there is some observer for whom these two events occur at the same time.  For that observer ${\sqrt { \Delta {s^2}}}$ is the spatial distance between these two points.  

Now let's move on from the simple case of motion along a line to the more general case of motion in three dimensional space.  Then an event has four coordinates $(t,x,y,z)$, and the spacetime interval is given by
\be
\Delta {s^2} = -\Delta {t^2} + \Delta {x^2} + \Delta {y^2} + \Delta {z^2} \; \; \; .
\label{interval2}
\ee
We now see that not only are boosts like rotations, but together boosts and rotations form the family of transformations (called Lorentz transformations) that leave the spacetime interval invariant.  

It is not only events that have four components, but also vectors since a boost that mixes up the $t$ and $x$ coordinates will also mix up the $x$ component of a vector with something that we will need to think of as the $t$ component of a four dimensional vector.  Note that the notation $(t,x,y,z)$ for a point in spacetime is somewhat cumbersome, but so is the notation $(x,y,z)$ for a point in space.  For that reason the standard vector notation has $\vec x$ stand for $(x,y,z)$ but we sometime also use $x^i$ where $i$ takes on the values 1, 2, and 3, and we have ${x^1}=x, \, {x^2}=y, \, {x^3}=z$.  In the spirit of this second notation, the standard special relativity notation for a point in spacetime is $x^\alpha$ where $\alpha$ takes on the values 0, 1, 2, and 3 and where ${x^0}=t, \, {x^1}=x, \, {x^2}=y, \, {x^3}=z$.  Similarly, we denote a four dimensional vector as $V^\alpha$.  

The spacetime interval can be thought of as a four dimensional dot product of $\Delta {x^\alpha}$ with itself.  However, this four dimensional dot product has one minus sign in addition to the familiar 3 plus signs.  One way to write this dot product is to introduce the quantities 
$\eta_{\alpha \beta}$ with ${\eta _{00}}=-1, \, {\eta _{11}}={\eta _{22}}={\eta _{33}}=1$ with all other components of $\eta_{\alpha \beta}$ equal to zero.  The spacetime interval can be written as 
\be
\Delta {s^2} = {\sum _{\alpha \beta}} {\eta _{\alpha \beta}} \Delta {x^\alpha}
\Delta {x^\beta} \; \; \; .
\label{dotproduct1}
\ee
This notation is somewhat cumbersome, so we will introduce two standard special relativity notations.  The first is the Einstein summation convention, which states that an index that is repeated in the upper and lower positions is summed over.  This allows to write eqn. (\ref{dotproduct1}) as
\be
\Delta {s^2} =  {\eta _{\alpha \beta}} \Delta {x^\alpha}
\Delta {x^\beta} \; \; \; .
\label{dotproduct2}
\ee
The second notation is that of lowering and raising indices by which for any vector $V^\alpha$ we have that $V_\alpha$ is a shorthand for ${\eta _{\alpha \beta}}{V^\beta}$. It then follows that we can raise indices with the inverse of $\eta _{\alpha \beta}$.  This inverse is denoted 
$\eta ^{\alpha \beta}$ and has components ${\eta ^{00}}= -1, \, {\eta ^{11}} = {\eta ^{22}} = {\eta ^{33}} = 1$ with all other components zero.  For any vector ${V^\alpha}$ we have
${V^\alpha}={\eta ^{\alpha \beta}}{V_\beta}$.  Due to the form of $\eta _{\alpha \beta}$, it follows that the rule for either raising or lowering is that spatial components are unchanged, while the time component is multiplied by minus one.  This allows us to write eqn. (\ref{dotproduct2}) as 
\be
\Delta {s^2} =   \Delta {x_\alpha}
\Delta {x^\alpha} \; \; \; .
\label{dotproduct3}
\ee
More generally, the dot product between any two vectors $A^\alpha$ and $B^\alpha$ is 
\be
{A^\alpha}{B_\alpha} = - {A^0}{B^0} + {A^1}{B^1} + {A^2}{B^2} + {A^3}{B^3} \; \; \; .
\ee

Special relativity is not simply some footnote to Newtonian mechanics: rather it demands that all the equations of physics be rewritten in terms of its basic structure: spacetime, four dimensional vectors, and the spacetime interval.  Newtonian mechanics begins with the notion that the motion of an object is to be written in terms of ${\vec x}(t)$, the spatial position as a function of time.  Instead we should write ${x^\alpha}(\tau)$ where $\tau$ is the proper time elapsed on a clock carried by the moving object.  So far we have only talked about proper time of an inertial observer.  But if we are to write ${x^\alpha}(\tau)$ we will also need an expression for proper time of an object undergoing acceleration.  

It is sometimes thought by retired engineers that one can't treat acceleration within special relativity, and for that one has to use general relativity.  But as we will see, that would be like saying that one can't treat curves in Euclidean geometry and would need to use non-Euclidean geometry.  (Which certainly would have been news to Euclid and the other ancient greek geometers who spent a lot of time thinking about circles).  Recall how one calculates the length of a curve.  The basic idea is to use the infinitesimal version of eqn. (\ref{length})
\be
d {\ell ^2} = d {x^2} + d {y^2} \; \; \; .
\label{length2}
\ee
and then think of the curve as a large number of tiny, approximately straight, segments.  The length of the curve is then
\be
L = \int d \ell = \int {\sqrt {d {x^2} + d {y^2}}} = \int dx {\sqrt {1 + {{\left ( {\frac {dy} {dx}} \right ) }^2}}} \; \; \; .
\ee
(where these are actually definite integrals with the limits of integration corresponding to the begining and end of the curve).

Now let's do the same with proper time.  The infinitesimal version of eqn. (\ref{interval2}) is
\be
d {\tau ^2} = d {t^2} - d {x^2} - d {y^2} - d {z^2} \; \; \; .
\label{length3}
\ee
We think of the curve in spacetime representing the history of the motion of the object as a large number of approximately straight segments.  The proper time of the curve is then
\bea
\tau &=& \int d \tau = \int {\sqrt {d {t^2} - d {x^2} - d {y^2} - d {z^2}}} 
\nonumber
\\
&=& \int dt {\sqrt {1 - {{\left ( {\frac {dx} {dt}} \right ) }^2} - {{\left ( {\frac {dy} {dt}} \right ) }^2}
- {{\left ( {\frac {dz} {dt}} \right ) }^2}}} 
= \int dt {\sqrt {1 - {v^2}}} \; \; \; .
\label{tau}
\eea

From here, things proceed in analogy to introductory physics: given the ``position'' $x^\alpha$ as a function of ``time'' $\tau$, we define the four-velocity $u^\alpha$ by 
${u^\alpha} \equiv d {x^\alpha}/d\tau$ and the four-acceleration $a^\alpha$ by
${a^\alpha} \equiv d {u^\alpha}/d\tau$.  The four-momentum $p^\alpha$ is defined by
${p^\alpha} \equiv m {u^\alpha}$ where $m$ is the mass of the object. 

We now look more closely at each of these quantities.  Since for an ordinary three dimensional vector $\vec V$ is an abreviation for $({V^x},{V^y},{V^z})$, we can write a four dimensional vector as ${V^\alpha}=({V^0},{\vec V})$.  Thus finding $u^\alpha$ means finding 
$dt/d\tau$ and $d{\vec x}/d\tau$.  Differentiating eqn. (\ref{tau}) with respect to $t$ we find $d\tau/dt={\sqrt {1-{v^2}}}$ and therefore $dt/d\tau =\gamma$.  We then find that $d{\vec x}/d\tau = (d{\vec x}/dt)(dt/d\tau)=\gamma {\vec v}$.  Thus we have
\be
{u^\alpha}=\gamma (1,{\vec v}) \; \; \; .
\label{velocity}
\ee

Note that $\gamma$ is a function of $\vec v$, so even though $u^\alpha$ has four components, only three of them are independent.  This suggests that the four components of $u^\alpha$ satisfy one algebraic relation, and indeed from eqn. (\ref{velocity}) we find
\be
{u^\alpha}{u_\alpha} = {\gamma ^2} ( -1 + {v^2}) = -1 \; \; \; .
\label{unitvelocity}
\ee
Differentiating eqn. (\ref{unitvelocity}) with respect to $\tau$ we find
\be
0 = {\frac d {d\tau}} ( {u^\alpha}{u_\alpha} ) = {u^\alpha}{a_\alpha} + {a^\alpha}{u_\alpha}
= 2 {a^\alpha}{u_\alpha} \; \; \; .
\label{accelerationconstraint}
\ee
Recall that the most important equation in Newtonian mechanics is ${\vec F} = m {\vec a}$.  We would like to have a relativistic version of this equation of the form ${f^\alpha} = m {a^\alpha}$.  But while we can write down almost any force law we like in Newtonian mechanics, eqn. (\ref{accelerationconstraint}) tells us that any relativistic force law must somehow satisfy ${f^\alpha}{u_\alpha}=0$.  We will return to (and expand upon) this point in section \ref{electrodynamics}.  

The spatial components of the four-momentum $p^\alpha$ are $\gamma m {\vec v}$.  For the Newtonian regime where $v$ is small these components are approximately equal to $m{\vec v}$ which is the standard Newtonian momentum, so whatever the velocity we continue to think of these components as momentum.  The time component is $\gamma m$, and as we will see shortly this is actually the relativistic energy.  This sounds like a major departure from Newtonian physics where the expression for the energy is $E=(1/2)m{v^2}$, but in a moment we will see that it isn't.  First note that though we have adopted units where $c=1$, we can always ``put the factors of $c$ back'' in any of our formulas by using dimensional analysis: that is using the fact that in ordinary units each quantity must have the appropriate units.  We then find that the relativistic expression for energy is 
\be
E = {\frac {m{c^2}} {\sqrt {1 - {\frac {v^2} {c^2}}}}} \; \; \; .
\label{energy}
\ee
For $v=0$ we recover Einstein's famous formula $E=m{c^2}$.  For $v$ small compared to $c$ ({\it i.e.} the regime in which Newtonian physics holds) we have
\be
{\frac 1 {\sqrt {1 - {\frac {v^2} {c^2}}}}} \approx 1 + {\frac {v^2} {2{c^2}}} \; \; \; ,
\ee
and therefore
\be
E \approx m{c^2} \left ( 1 + {\frac {v^2} {2{c^2}}} \right ) = m {c^2} + {\textstyle {\frac 1 2}} m {v^2} \; \; \; .
\ee
This is just the standard Newtonian result plus a constant $m{c^2}$.  But this is actually consistent with Newtonian physics which only talks about energy differences and thus says that $(1/2) m {v^2} $ is the difference between the energy at velocity $v$ and the energy at velocity zero.  It is only a convention in Newtonian physics that sets that constant to zero.

Eqn. (\ref{energy}) gives the physical reason why nothing can travel faster than light and why all objects travel slower than light: in this formula, the energy goes to infinity as the speed goes to $c$.  Solving eqn. (\ref{energy}) for $v$ we find
\be
v = c {\sqrt {1 - {{\left ( {\frac {m{c^2}} E} \right ) }^2}}} \; \; \; .
\ee
This formula has been very well tested in particle accelerators.  As we put more and more energy into a particle it gets ever closer to the speed of light, but never reaches that speed.

\section{Electrodynamics in special relativity}
\label{electrodynamics}

The usual textbook method of presenting electrodynamics is to use the same time and space formalism as in Newtonian mechanics.  This method is either used exclusively throughout the book, or at best there is one chapter in which electrodynamics is presented in relativistic notation.  However, there are notable exceptions\cite{ll2} in which electrodynamics is presented in relativistic notation throughout.  

It is not widely appreciated that electrodynamics is actually simpler and makes more sense when done from the begining in relativistic notation.  Let's begin by considering the equation of motion of an object with mass $m$ and charge $q$ under the Lorentz force law.
\be
m {\vec a} = q ({\vec E} + {\vec v} \times {\vec B} ) \; \; \; ,
\label{lorentzforce1}
\ee  
where $\vec E$ is the electric field and $\vec B$ is the magnetic field.  Somehow there are two different fields producing two different forces, one velocity dependent and one not, with the velocity dependent force for some reason involving the cross product.  That looks like an odd and complicated force law.

Now let's consider the same force law in special relativity.  
\be
m {a_\alpha} = q {F_{\alpha \beta}}{u^\beta} \; \; \; ,
\label{lorentzforce2}
\ee
where $F_{\alpha \beta}$ is antisymmetric: that is ${F_{\beta \alpha}}=-{F_{\alpha \beta}}$.  Because of this antisymmetry property, it immediately follows that for any $u^\alpha$ we have
${F_{\alpha \beta}}{u^\alpha}{u^\beta}=0$ and therefore from eqn. (\ref{lorentzforce2}) that 
${u^\alpha}{a_\alpha}=0$.  Thus the Lorentz force law of eqn. (\ref{lorentzforce2}) is perhaps the simplest force law that one can write down that satisfies the condition of eqn. (\ref{accelerationconstraint}) 
that is required for a relativistic force law.  

Now let's see how we recover the Newtonian version of the equation of motion, eqn. (\ref{lorentzforce1}) for velocities small compared to $c$.  In this case we have $\gamma \approx 1$ and therefore ${u^\alpha} \approx (1,{\vec v})$.  Equation (\ref{lorentzforce1}) then becomes 
\be
m {a_i} = q ({F_{i0}}+{F_{ij}}{v^j}) \; \; \; .
\label{lorentzforce3}
\ee
Comparing eqns. (\ref{lorentzforce1}) and (\ref{lorentzforce3}) it is clear that 
\be
{F_{i0}}={E_i} \; \; \; ,
\label{electric}
\ee
or in other words that the time-space components of $F_{\alpha \beta}$ are the electric field.  It also seems that somehow the space-space components of $F_{\alpha \beta}$ are the components of the magnetic field, but it is not entirely clear how since eqn. (\ref{lorentzforce1}) contains the cross product while eqn. (\ref{lorentzforce3}) does not.  To explain this point we note that one formula for the cross product involves the epsilon symbol.  For any two vectors $\vec A$ and $\vec B$
\be
{{({\vec A} \times {\vec B})}_i}={\epsilon _{ijk}}{A^j}{B^k} \; \; \; .
\ee
Here the epsilon symbol satisfies the property that $\epsilon_{ijk}$ is zero if any two indices are the same, that ${\epsilon_{123}}=1$ and that switching the position of any two indices results in multiplying the symbol by minus one.  Thus we see that $ {{({\vec v} \times {\vec B})}_i} = {F_{ij}}{v^j}$ provided that 
\be
{F_{ij}}={\epsilon _{ijk}}{B^k} \; \; \; .
\label{magnetic}
\ee

If the calculations of the previous paragraph seemed complicated and messy, note that none of that complication is a property of special relativity, but only of the expression of special relativity in Newtonian terms.  Indeed that is a major theme of this paper: that special relativity viewed on its own terms is a sensible, coherent, and ultimately simple subject.  The complications come from expressing the truths of special relativity in Newtonian terms, while the paradoxes come from trying to understand special relativity while still insisting that all one's Newtonian intuitions are correct.

To illustrate the point that special relativity on its own terms is simple, we will do two calculations: (a) motion of a charge in a constant electric field and (b) motion of a charge in a constant magnetic field.

Consider a constant electric field of magnitude $E$ whose direction we will choose to be the $x$ axis.  Then the only nonzero components of $F_{\alpha \beta}$ are ${F_{10}}=E$ and ${F_{01}}=-E$.  We will assume that at time zero the charge is at rest at the origin.  From eqn. (\ref{lorentzforce2}) we then find
\bea
m {\frac {d{u_0}} {d\tau}} &=& - q E {u^1} \; \; \; ,
\label{motion1a}
\\
m {\frac {d{u_1}} {d\tau}} &=&  q E {u^0} \; \; \; .
\label{motion1b}
\eea
Now define the constant $a$ to be $a \equiv qE/m$.  We will later show that $a$ is the invariant magnitude of the acceleration of the charge, but for now it is just a constant.  Using the fact that ${u^0}=-{u_0}$ and ${u^1}={u_1}$ we find
\bea
{\frac {d{u^0}} {d\tau}} &=& a {u^1} \; \; \; ,
\label{motion2a}
\\
{\frac {d{u^1}} {d\tau}} &=&  a {u^0} \; \; \; .
\label{motion2b}
\eea
Taking the derivative of eqn. (\ref{motion2b}) and using eqn. (\ref{motion2a}) we find
\be
{\frac {{d^2}{u^1}} {d {\tau^2}}} = {a^2} {u^1} \; \; \; ,
\ee
whose solution is
\be
{u^1} = \sinh a \tau \; \; \; .
\label{motion3}
\ee
It then follows from eqn. (\ref{motion2a}) that
\be
{u^0} = \cosh a \tau \; \; \; .
\label{motion4}
\ee
Finally integrating eqns. (\ref{motion3}) and (\ref{motion4}) we obtain
\bea
t &=& {\frac 1 a} \sinh a \tau \; \; \; ,
\label{motion5a}
\\
x &=& {\frac 1 a} (\cosh a \tau - 1) \; \; \; .
\label{motion5b}
\eea
Though eqns. (\ref{motion5a}) and (\ref{motion5b}) express both $t$ and $x$ as functions of $\tau$, we can use them together to express $x$ as a function of $t$ which yields
\be
x = {\frac 1 a} \left ( {\sqrt {1+{a^2}{t^2}}} - 1 \right ) \; \; \; .
\label{motion6}
\ee
Note that for $a t \ll 1$ we have $x \approx (1/2) a {t^2}$ as we would expect from Newtonian physics, while for $a t \gg 1$ we have $x \approx t$ as we would expect for an object traveling at nearly the speed of light.  Finally we justify the interpretation of $a$ as an invariant magnitude of acceleration.  Differentiating eqns. (\ref{motion4}) and (\ref{motion3}) we obtain
\bea
{a^0} = a \sinh a \tau \; \; \; ,
\\
{a^1} = a \cosh a \tau \; \; \; ,
\eea
from which we find
\be
{\sqrt {{a^\alpha}{a_\alpha}}} = {\sqrt {{{({a^1})}^2} - {{({a^0})}^2}}} 
= {\sqrt {{a^2} {\cosh^2} a \tau - {a^2} {\sinh^2} a \tau}} = a \; \; \; .
\ee

We now turn to motion in a constant magnetic field with magnitude $B$ and whose direction we will take to be the $z$ axis.  Then the only nonzero components of $F_{\alpha \beta}$ are ${F_{12}}=B$ and ${F_{21}}=-B$.  We will assume that at time zero the velocity of the charge is in the $y$ direction.  From eqn. (\ref{lorentzforce2}) we then find
\bea
m {\frac {d{u_1}} {d\tau}} &=&  q B {u^2} \; \; \; ,
\label{bmotion1a}
\\
m {\frac {d{u_2}} {d\tau}} &=&  -q B {u^1} \; \; \; .
\label{bmotion1b}
\eea
Now define the constant $\omega$ to be $\omega \equiv qB/m$.  This constant $\omega$ is called the cyclotron frequency for reasons that will become clear later.  Using the fact that ${u^1}= {u_1}$ and ${u^2}={u_2}$ we find
\bea
{\frac {d{u^1}} {d\tau}} &=& \omega {u^2} \; \; \; ,
\label{bmotion2a}
\\
{\frac {d{u^2}} {d\tau}} &=&  -\omega {u^1} \; \; \; .
\label{bmotion2b}
\eea
Taking the derivative of eqn. (\ref{bmotion2a}) and using eqn. (\ref{bmotion2b}) we find
\be
{\frac {{d^2}{u^1}} {d {\tau^2}}} = - {\omega^2} {u^1} \; \; \; ,
\ee
whose solution is
\be
{u^1} = - k \sin \omega \tau \; \; \; ,
\label{bmotion3}
\ee
for some constant $k$ whose physical meaning will become clear once we have expressions for $x$ and $y$.
It then follows from eqn. (\ref{bmotion2a}) that
\be
{u^2} = k \cos \omega \tau \; \; \; .
\label{bmotion4}
\ee
Finally integrating eqns. (\ref{bmotion3}) and (\ref{bmotion4}) we obtain
\bea
x &=& {\frac k \omega} \cos \omega \tau \; \; \; ,
\label{bmotion5a}
\\
y &=& {\frac k \omega} \sin \omega \tau \; \; \; .
\label{bmotion5b}
\eea
Equations (\ref{bmotion5a}) and (\ref{bmotion5b}) describe circular motion, and we have chosen the constants of integration so that the center of the circle is at the origin.  The radius of the circle is $R=k/\omega$ which means that our original constant of integration $k$ can be written as $k=R\omega$.  Thus eqns. (\ref{bmotion3}-\ref{bmotion5b}) become
\bea
{u^1} &=& - R\omega \sin \omega \tau \; \; \; ,
\label{bmotion6a}
\\
{u^2} &=& R\omega \cos \omega \tau \; \; \; ,
\label{bmotion6b}
\\
x &=& R \cos \omega \tau \; \; \; .
\label{bmotion6c}
\\
y &=& R \sin \omega \tau \; \; \; .
\label{bmotion6d}
\eea

We now turn to the other components of $u^\alpha$.  Equation (\ref{lorentzforce2}) says that $u^0$ and $u^3$ are constants.  For simplicity we chose initial condition with ${u^3}=0$, but if we had chosen some other initial conditions, this would leave eqns. (\ref{bmotion6a}-\ref{bmotion6d}) unchanged and simply add to them the equations ${u^3}={c_0}$ and $z={c_0}\tau$ with $c_0$ a constant.  That is along with circular motion in the plane we would have linear motion in the $z$ direction: the trajectory would be a helix.

As for $u^0$, its value can be found by the algebraic condition that ${u^\alpha}{u_\alpha}=-1$.
\bea
-1 &=& {u^\alpha}{u_\alpha} = - {{({u^0})}^2} + {{({u^1})}^2} + {{({u^2})}^2} 
\nonumber
\\
&=& - {{({u^0})}^2} + {{(- R\omega \sin \omega \tau )}^2} + {{(R\omega \cos \omega \tau )}^2} 
= - {{({u^0})}^2} + {R^2}{\omega^2} \; \; \; ,
\eea
which leads to 
\bea
{u^0}&=&{\sqrt {1 + {R^2}{\omega^2}}} \; \; \; ,
\label{magneticu0}
\\
t&=&\tau {\sqrt {1 + {R^2}{\omega^2}}} \; \; \; .
\label{magnetict}
\eea

We now turn to the origin of the name ``cyclotron frequency'' for $\omega$.  Much of the study of fundamental physics is explored through the behavior of particles of high energy.  To produce such particles, we need to acclerate them.  From the first example of this section, it is clear that we could do this using electric fields.  But to produce particles of extremely high energy we would need either very large electric fields or to maintain moderately large electric fields over large distances.  An alternative method uses the magnetic field calculation we have just done.  First consider the case $R\omega \ll 1$ which corresponds to the speed of the particle being small compared to $c$.  Then $t \approx \tau$.  Thus the particle goes around the circle once in a time $T=2 \pi /\omega$ no matter what value $R$ has.  Now suppose we have a microwave generator with angular frequency exactly $\omega$.  Since the voltage is alternating, we can hook it to electrodes that will give the particle a small kick in the same direction that it is moving twice per cycle.  The result is that though the period of the particle stays the same, each cycle its speed gets a little larger and so does its radius $R$.  We can get to large energies without using large voltages since the energy is delivered one small kick at a time.  The machine that does this is called a cyclotron: hence the name ``cyclotron frequency.''

But special relativity had a nasty surprise for the designers of the cyclotron: once the speed becomes comparable to the speed of light, we need to use a more complicated formula for the time $T$ it takes the particle to go around the circle.  The amount of proper time that it takes to go around the circle is still $2\pi/\omega$. So from eqn. (\ref{magnetict}) we find
\be
T = {\frac {2 \pi {\sqrt {1 + {R^2}{\omega^2}}}} \omega} \; \; \; .
\ee 
Thus as the particle's energy increased it eventually became out of synch with the microwave generator.  At high speeds the old design didn't work, so a new design was needed.  The new design is called a ``snychrotron'' (shortened from the more clumsy term ``snychro-cyclotron'')
and is the design used in the Large Hadron Collider.  Here magnetic fields still keep the particles in circular motion, though the circle is an enormous ring of pipes with each pipe having its own magnet.  The acceleration is no longer done by a microwave generator of a fixed frequency, but instead by a carefully timed series of pulses delivered at exactly the right time to exactly the right place in the ring.

We now turn to the form of Maxwell's equations in both the usual notation and in the notation of special relativity.  The physical content of Maxwell's equations is that (1) charges make electric fields, (2) moving charges make magnetic fields, (3) changing magnetic fields make electric fields, and (4) changing electric fields make magnetic fields.  This content is contained in the following four equations:
\bea
{\vec \nabla} \cdot {\vec E} &=& 4 \pi \rho \; \; \; ,
\label{maxwell1}
\\
{\vec \nabla} \cdot {\vec B} &=& 0 \; \; \; ,
\label{maxwell2}
\\
{\vec \nabla} \times {\vec E} + {\frac 1 c} {\frac {\partial {\vec B}} {\partial t}} &=& 0
\; \; \; ,
\label{maxwell3}
\\
{\vec \nabla} \times {\vec B} &=& {\frac {4\pi} c} {\vec j} +  {\frac 1 c} {\frac {\partial {\vec E}} {\partial t}} \; \; \; .
\label{maxwell4}
\eea
Here $\rho$ is the charge density and $\vec j$ is the current density.  

In the notation of special relativity there are only two Maxwell equations:
\bea
{\partial _\alpha}{F_{\beta \gamma}} + {\partial _\beta}{F_{\gamma \alpha}} +
{\partial _\gamma}{F_{\alpha \beta}} &=& 0 \; \; \; ,
\label{srmaxwell1}
\\
{\partial _\alpha}{F^{\alpha \beta}} = - 4 \pi {j^\beta} \; \; \; .
\label{srmaxwell2}
\eea
Here ${j^\alpha}=(\rho,{\vec j})$ is the current density four-vector, and $\partial _\alpha$ means partial derivative with respect to $x^\alpha$.  Here eqn. (\ref{srmaxwell1}) has the content of eqns. (\ref{maxwell2}) and (\ref{maxwell3}), while eqn. (\ref{srmaxwell2}) has the content of eqns. (\ref{maxwell1}) and (\ref{maxwell4}).  

It is clearly simpler to have two equations instead of four.  Nonetheless, many people encountering relativistic electrodynamics are offput by the appearence of the antisymmetric tensor $F_{\alpha \beta}$ when in other areas of physics we tend to use only scalars and vectors.  So it is worthwhile to look at the use of potentials in Maxwell's equations, in both the standard notation and the relativistic notation.  

In the standard notation, one notices that due to the identity ${\vec \nabla}\cdot ( {\vec \nabla} \times {\vec V}) = 0$ for any vector field $\vec V$, it follows that if the magnetic field ${\vec B}$ takes the form 
\be
{\vec B} = {\vec \nabla} \times {\vec A} \; \; \; ,
\label{magpotential}
\ee
for some vector potential $\vec A$ then eqn. (\ref{maxwell2}) is automatically satisfied. Equation (\ref{maxwell3}) then becomes
\be
{\vec \nabla} \times \left ( {\vec E} + {\frac 1 c} {\frac {\partial {\vec A}} {\partial t}} \right ) = 0 \; \; \; .
\ee
Therefore there is a scalar $\Phi$ such that 
\be
{\vec E} + {\frac 1 c} {\frac {\partial {\vec A}} {\partial t}}  = - {\vec \nabla}\Phi \; \; \; ,
\ee
and therefore the electric field is given by
\be
{\vec E} =  - {\vec \nabla}\Phi - {\frac 1 c} {\frac {\partial {\vec A}} {\partial t}} \; \; \; .
\label{elecpotential}
\ee

In the notation of special relativity all the reasoning of the previous paragraph is contained in the statement that if there is a vector field $A_\alpha$ such that
\be
{F_{\alpha \beta}} = {\partial _\alpha}{A_\beta} - {\partial _\beta} {A_\alpha} \; \; \; ,
\label{srpotential}
\ee
then eqn. (\ref{srmaxwell1}) is automatically satisfied.

In the standard notation one notes that the potentials $\Phi$ and $\vec A$ for a given electric and magnetic field are not unique: if one takes a given vector $\vec A$ and produces a new $\vec A$ by ${\vec A} \to {\vec A} + {\vec \nabla} \chi$ for some scalar $\chi$ then $\vec B$ given by
eqn. (\ref{magpotential}) is unchanged.  Furthermore if one also changes $\Phi$ by 
$\Phi \to \Phi - (1/c)\partial \chi/\partial t$ using {\emph {the same}} $\chi$ then $\vec E$ given by eqn. (\ref{elecpotential}) is also unchanged.

In the notation of special relativity all the reasoning of the previous paragraph is that if one changes $A_\alpha$ by ${A_\alpha} \to {A_\alpha} + {\partial _\alpha} \chi$ for any scalar $\chi$ then $F_{\alpha \beta}$ given by eqn. (\ref{srpotential}) is unchanged.

Now we choose a particular $\chi$ so that $\Phi$ and $\vec A$ satisfy
\be
{\vec \nabla}\cdot {\vec A} + {\frac 1 c} {\frac {\partial \Phi} {\partial t}} = 0 \; \; \; .
\ee
Then using eqn. (\ref{elecpotential}) in eqn. (\ref{maxwell1}) we find that $\Phi$ satisfies the equation
\be
{\nabla ^2} \Phi - {\frac 1 {c^2}} {\frac {{\partial ^2}\Phi} {\partial {t^2}}} = - 4 \pi \rho
\; \; \; .
\ee
Furthermore using eqn. (\ref{magpotential}) in eqn. (\ref{maxwell4}) we find that $\vec A$ satisfies the equation
\be
{\nabla ^2} {\vec A} - {\frac 1 {c^2}} {\frac {{\partial ^2}{\vec A}} {\partial {t^2}}} = - 
{\frac {4 \pi {\vec j}} c} \; \; \; .
\ee

In the notation of special relativity all the reasoning of the previous paragraph becomes the following: by a particular choice of $\chi$ we can arrange for $A^\alpha$ to satisfy the condition ${\partial _\alpha}{A^\alpha}=0$.  Then eqn. (\ref{srmaxwell2}) becomes
\be
{\partial _\alpha}{\partial ^\alpha}{A^\beta} = - 4 \pi {j^\beta} \; \; \; .
\ee

Thus, in summary we see that electrodynamics is much simpler and more straightforward in the language of special relativity.

\section{Clocks, light, radar, and transponders}
\label{clocks1}

We now come to the operational meaning of the boosts.  As was appropriate for 1905, Einstein liked to state things in terms of networks of observers with clocks and rulers, one on the ground and one on a moving train, with the clocks synchronized using light signals.  Over 120 years later we will find it easier to talk about individual observers each with their own clocks and with the ability to use radar.  

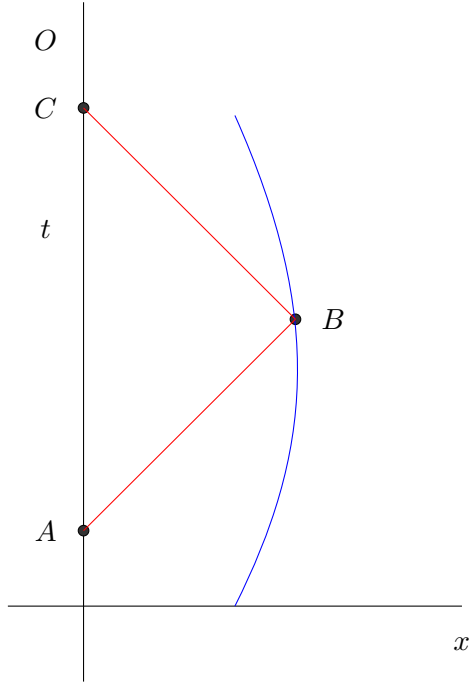
\begin{figure}
\centering
\begin{tikzpicture}
\draw (-1,0) -- (5,0);
\draw (0,-1) -- (0,8);
\node at (5,-0.5) {$x$};
\node at (-0.5,5) {$t$};
\node at (-0.5,7.5) {$O$};
\node at (-0.5,1) {$A$};
\node at (3.3,3.8) {$B$};
\node at (-0.5,6.6) {$C$};
\draw[fill=black!80] (0,1) circle (0.07);
\draw[fill=black!80] (2.8,3.8) circle (0.07);
\draw[fill=black!80] (0,6.6) circle (0.07);
\draw[blue] (2,0) .. controls (3,2) and (3.2,3.8) .. (2,6.5);
\draw[red] (0,1) -- (2.8,3.8);
\draw[red] (2.8,3.8) -- (0,6.6);
\end{tikzpicture}
\caption{An observer $O$ ($t$ axis) measures a moving object (blue curve) using radar (red lines)}
\label{radar}
\end{figure}

Recall that radar uses the round trip time of a radio wave ({\it i.e.} low frequency light) pulse to deduce the distance of an object.  Note that since this is done with moving objects, we are deducing the distance of the object at a particular time.  As with many things, we will find it useful to draw a spacetime diagram of the operation of radar, see figure (\ref{radar}).
A spacetime diagram is a picture of the history of the observers, objects, and light rays that we wish to analyze.  Here space is the $x$ axis and time is the $y$ axis.  Since we have chosen units where $c=1$, this means that the history of any light ray is a line at an angle of 45 degrees to the vertical.  In figure (\ref{radar}) the $t$ axis is the history (also called the worldline) of the observer $O$.  In other words, this observer remains at $x=0$ at all times.  The blue curve is the worldline of the moving object that $O$ is measuring with radar.  Points on the diagram are called events, with each event being a single point of space at a single moment of time.  Event $A$ is when the observer sends out his outgoing radar pulse, event $B$ is when that pulse hits and reflects off the object, and event $C$ is when the return pulse is received by the observer.  The history of the outgoing light pulse is the red line from $A$ to $B$, while the history of the reflected light pulse is the red line from $B$ to $C$.  

Let $\tau _A$ be the time on observer $O$'s clock at event $A$ and let $\tau _C$ be the time on observer $O$'s clock at event $C$.  Then famously radar deduces the distance of the object as 
$({\tau _C}-{\tau _A})/2$.  More precisely, since the object is moving, radar deduces that the position of event $B$ (the object when it reflected the radar pulse) is $({\tau _C}-{\tau _A})/2$.  Similarly $O$ can deduce that the time of event $B$ is the average of the time when the outgoing pulse was emitted and the time when the reflected pulse is received.  That is, the time that observer $O$ measures for event $B$ is $({\tau _C}+{\tau _A})/2$.

We will take the point of view that coordinate systems are essential tools for doing calculations, but that physical properties are to be read off from the behavior of clocks and light.  (This point of view is helpful for special relativity, but absolutely essential for general relativity).  So we suppose that in $(t,x)$ coordinates event $B$ is given by $({t_B},{x_B})$ and ask what time and position the radar measurements of $O$ assign to that event.  The coordinates of event $A$ are $({t_A},0)$ but also since event $A$ is on a line of slope 1 that contains event $B$ we find that ${t_A}={t_B}-{x_B}$.  Similarly, the coordinates of event $C$ are $({t_C},0)$ but also since event $C$ is on a line of slope -1 that contains event $B$ we find that ${t_C}={t_B}+{x_B}$.

Clocks measure proper time $\tau$ where $\Delta {\tau^2}=\Delta {t^2} - \Delta {x^2}$.  Since observer $O$ is at $x=0$ it follows that $\Delta x = 0$ for his clock and therefore $\Delta \tau = \Delta t$.  Setting the clock to zero at the origin (that is, the point $(0,0)$) we find that
$\tau = t$ and therefore that ${\tau _A}={t_A}$ and ${\tau_C}={t_C}$.  It then follows that the time measured by observer $O$ for event $B$ is 
\be
{\frac {{\tau _C}+{\tau _A}} 2} = {\frac {{t_C}+{t_A}} 2} = {\frac {({t_B}+{x_B})+({t_B}-{x_B})} 2} = {t_B} \; \; \; .
\ee
Similarly it follows that the position measured 
by observer $O$ for event $B$ is 
\be
{\frac {{\tau _C}-{\tau _A}} 2} = {\frac {{t_C}-{t_A}} 2} = {\frac {({t_B}+{x_B})-({t_B}-{x_B})} 2} = {x_B} \; \; \; .
\ee
Since $B$ can be any event, we have deduced that for an event with coordinates $(t,x)$ the time measured by $O$ is $t$ and the position measured by $O$ is $x$.

\begin{figure}
\centering
\begin{tikzpicture}
\draw (-1,0) -- (5,0);
\draw (0,-1) -- (0,8);
\node at (5,-0.5) {$x$};
\node at (-0.5,5) {$t$};
\node at (-0.5,7.5) {$O$};
\node at (1.9,7.5) {$O'$};
\node at (-0.5,1) {$A$};
\node at (3.3,3.8) {$B$};
\node at (-0.5,6.6) {$C$};
\node at (0.93,1.43) {$D$};
\node at (1.02,5.08) {$E$};
\draw[fill=black!80] (0,1) circle (0.07);
\draw[fill=black!80] (2.8,3.8) circle (0.07);
\draw[fill=black!80] (0,6.6) circle (0.07);
\draw[fill=black!80] (0.43,1.43) circle (0.07);
\draw[fill=black!80] (1.52,5.08) circle (0.07);
\draw[blue] (2,0) .. controls (3,2) and (3.2,3.8) .. (2,6.5);
\draw[red] (0,1) -- (2.8,3.8);
\draw[red] (2.8,3.8) -- (0,6.6);
\draw (-0.3,-1) -- (2.4,8);
\end{tikzpicture}
\caption{An observer $O$ ($t$ axis) and another observer $O'$ ($x=vt$) both measure a moving object (blue curve) using radar (red lines)}
\label{twoobservers}
\end{figure}
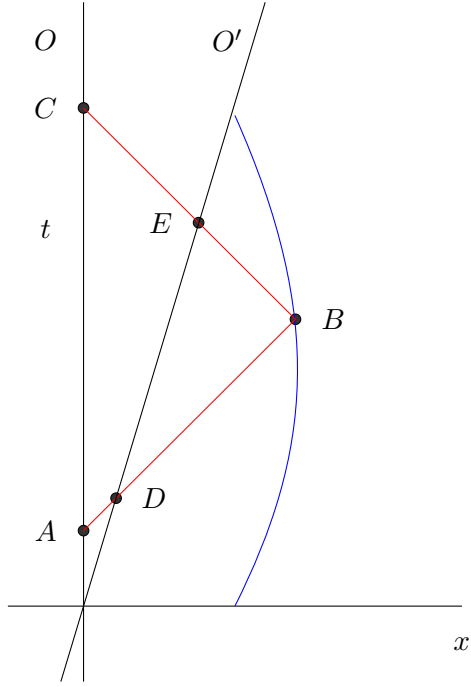

This might seem like an overly elaborate calculation for such a simple result, but now that we have practice doing this sort of calculation, we can do it again for a moving observer and thus derive the boost.  Consider figure (\ref{twoobservers}) which is just figure (\ref{radar}) with the worldline of a moving observer $O'$ added.  This observer is moving at velocity $v$ and thus his worldline is the line $x=vt$.  
Though we have the same red lines for the radar pulses, for the purpose of the measurement made by $O'$ we can consider the outgoing pulse as going from event $D$ to event $B$ and the reflected pulse as going from event $B$ to event $E$.  The time and space positions measured by $O'$ for event $B$ will then be $({{t'}_B},{{x'}_B})$ where ${{t'}_B}=({\tau_E}+{\tau_D})/2$
and ${{x'}_B}=({\tau_E}-{\tau_D})/2$.  Here $\tau_D$ is the time measured on the clock carried by $O'$ at event $D$ (and correspondingly for $\tau_E$).

We begin by working out the $(t,x)$ coordinates of the events $D$ and $E$.  Since event $D$ is both on the line $x=vt$ and on a line of slope 1 that contains event $B$ we find that its 
coordinates $({t_D},{x_D})$ are given by 
\be
{t_D} = {\frac {{t_B}-{x_B}} {1-v}} \; \; \; ,
\ee
and ${x_D}=v{t_D}$.  Similarly since event $E$ is both on the line $x=vt$ and on a line of slope -1 that contains event $E$ we find that its 
coordinates $({t_E},{x_E})$ are given by 
\be
{t_E} = {\frac {{t_B}+{x_B}} {1+v}} \; \; \; ,
\ee
and ${x_E}=v{t_E}$.  

Clocks measure proper time $\tau$ where $\Delta {\tau^2}=\Delta {t^2} - \Delta {x^2}$.  Since the worldline of observer $O$ is $x=vt$ it follows that $\Delta x = v \Delta t$ for his clock and therefore $\Delta \tau = {\gamma ^{-1}} \Delta t$.  Setting the clock to zero at the origin (that is, synchronizing the clocks of $O$ and $O'$ at the event where they meet) we find that
$\tau = {\gamma^{-1}} t$ and therefore that ${\tau _D}={\gamma^{-1}}{t_D}$ and ${\tau_E}={\gamma^{-1}}{t_E}$.  It then follows that the time measured by observer $O'$ for event $B$ is 
\bea
{{t'}_B} &=& {\frac {{\tau _E}+{\tau _D}} 2} = {\frac {{t_E}+{t_D}} {2\gamma}} = 
 {\frac {{t_B}+{x_B}} {2\gamma (1+v)}} + {\frac {{t_B}-{x_B}} {2\gamma (1-v)}}  
\nonumber
\\
&=& {\frac 1 {2 \gamma (1-{v^2})}} \left [ (1-v)({t_B}+{x_B}) + (1+v)({t_B}-{x_B}) \right ] 
= \gamma ({t_B} - v {x_B}) \; \; \; .
\eea
Similarly it follows that the position measured 
by observer $O'$ for event $B$ is 
\bea
{{x'}_B} &=& {\frac {{\tau _E}-{\tau _D}} 2} = {\frac {{t_E}-{t_D}} {2\gamma}} = 
 {\frac {{t_B}+{x_B}} {2\gamma (1+v)}} - {\frac {{t_B}-{x_B}} {2\gamma (1-v)}}  
\nonumber
\\
&=& {\frac 1 {2 \gamma (1-{v^2})}} \left [ (1-v)({t_B}+{x_B}) - (1+v)({t_B}-{x_B}) \right ] 
= \gamma ({x_B} - v {t_B}) \; \; \; .
\eea
Since $B$ can be any event, we have deduced that for an event with coordinates $(t,x)$ the  coordinates $({t'},{x'})$ measured by $O'$ are given by
\bea
{t'} = \gamma (t-vx) \; \; \; ,
\nonumber
\\
{x'} = \gamma (x-vt) \; \; \; .
\eea
Thus the boost is a direct consequence of the behavior of clocks and light.

From the boost we can immediately derive the phenomenon of time dilation: that each observer thinks that the other's clock is running slow.  First consider any point $(t,vt)$ on the worldline of $O'$.  Then $O$ says the time at this point is $t$ but the reading on the clock of $O'$ is ${t'}=\gamma (t-vx)=\gamma (t-{v^2}t)=t/\gamma$.  Thus $O$ says that the clock of $O'$ is running slow by a factor of $\gamma$.  Now consider any point $(t,0)$ on the worldline of $O$.  Then the reading on the clock of $O$ is $t$ but $O'$ says the time is ${t'}=\gamma(t-vx)=\gamma t$. Thus $O'$ says that the clock of $O$ is running slow by a factor of $\gamma$.  

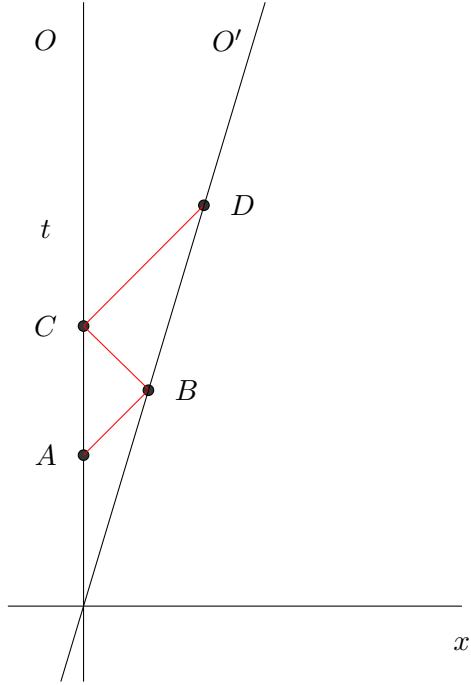
\begin{figure}
\centering
\begin{tikzpicture}
\draw (-1,0) -- (5,0);
\draw (0,-1) -- (0,8);
\node at (5,-0.5) {$x$};
\node at (-0.5,5) {$t$};
\node at (-0.5,7.5) {$O$};
\node at (1.9,7.5) {$O'$};
\node at (-0.5,2) {$A$};
\node at (1.36,2.86) {$B$};
\node at (-0.5,3.71) {$C$};
\node at (2.1,5.31) {$D$};
\draw[fill=black!80] (0,2) circle (0.07);
\draw[fill=black!80] (0.857,2.86) circle (0.07);
\draw[fill=black!80] (0,3.71) circle (0.07);
\draw[fill=black!80] (1.59,5.31) circle (0.07);
\draw[red] (0,2) -- (0.857,2.86);
\draw[red] (0.87,2.86) -- (0,3.71);
\draw[red] (0,3.71) -- (1.59,5.31);
\draw (-0.3,-1) -- (2.4,8);
\end{tikzpicture}
\caption{An observer $O$ ($t$ axis) and another observer $O'$ ($x=vt$) exchange signals}
\label{catch}
\end{figure}

Despite time dilation being a simple consequence of the boost, it still seems odd that each observer calls the other's clock slow.  To get a more operational feel for this phenomenon, we will consider a kind of game of catch between the two observers using radio waves and transponders as illustrated in figure (\ref{catch}).  
This game works as follows: at event $A$ observer $O$ sends a radio signal that contains the reading on his clock.  Each time $O'$ receives a signal from $O$ he immediately sends a signal back giving the reading on his clock.  Each time $O$ receives a signal from $O'$ he immediately sends a signal back giving the reading on his clock.  This results in signals sent from event $A$ to event $B$ to event $C$ to event $D$, and so on.  

At event $C$ observer $O$ has received the time $\tau _B$ that the clock of $O'$ read at event $B$.  But he also knows the time he would attribute to event $B$, namely $({\tau _A}+{\tau _C})/2$.  Thus he can compare these two times to see whether he should think of the clock of $O'$ as running slow, and if so by how much.  Similarly, at event $D$ observer $O'$ has received the time $\tau_C$ that the clock of $O$ read at event $C$.  But he also knows that he would attribute to the event $C$ the time $({\tau _B}+{\tau _D})/2$.  He can compare the two times to see whose clock is running slow.  

We now perform the calculations to do these comparisons.  Event $A$ has the coordinates $({t_A},0)$.  Since event $B$ with coordinates $({t_B},{x_B})$ is on a line of slope 1 that contains 
event $A$ and also on the line $x=vt$ we find that ${t_B}={t_A}/(1-v)$ and ${x_B}=v{t_B}$.  Event $C$ with coordinates $({t_C},0)$ is on a line with slope -1 that contains event $B$.  We therefore find
\be
{t_C}=(1+v){t_B}=  {\frac {1+v} {1-v}}  {t_A} \; \; \; .
\ee
Then applying to event $D$ with coordinates $({t_D},{x_D})$ the same reasoning that we used for event $B$ we find that
\be
{t_D} = {\frac {t_C} {1-v}} = {\frac {1+v} {{(1-v)}^2}}  {t_A} \; \; \; .
\ee
For the clock of $O$ we have $\tau =t$, while for the clock of $O'$ we have $\tau =t/\gamma$.  Thus we have
\bea
{\tau_A}&=&{t_A} \; \; \; ,
\nonumber
\\
{\tau_B}&=& {\frac 1 {\gamma (1-v)}} {t_A} \; \; \; ,
\nonumber
\\
{\tau_C}&=&  {\frac {1+v} {1-v}}  {t_A} \; \; \; ,
\nonumber
\\
{\tau_D}&=& {\frac {1+v} {\gamma {{(1-v)}^2}}}  {t_A} \; \; \; .
\eea

We then find 
\be
{\frac {{\tau_A}+{\tau_C}} 2} = {\frac {t_A} {2(1-v)}} \left [ (1-v)+(1+v) \right ] = 
{\frac {t_A} {1-v}} = \gamma {\tau_B} \; \; \; .
\ee
Thus the time attributed by observer $O$ to event $B$ is $\gamma $ multiplied by the time on the clock of $O'$ at event $B$.  Observer $O$ therefore concludes that the clock of $O'$ runs slow by a factor of $\gamma$.  

However we also have 
\be
{\frac {{\tau_B}+{\tau_D}} 2} = {\frac {t_A} {2\gamma {{(1-v)}^2}}} \left [ (1-v)+(1+v) \right ] = 
{\frac {t_A} {\gamma {{(1-v)}^2}}} = \gamma {\tau_C} \; \; \; .
\ee
Thus the time attributed by observer $O'$ to event $C$ is $\gamma $ multiplied by the time on the clock of $O$ at event $C$.  Observer $O'$ therefore concludes that the clock of $O$ runs slow by a factor of $\gamma$.

Therefore in operational terms $O$ has good reason to conclude that the clock of $O'$ runs slow {\emph {and}} $O'$ has good reason to conclude that the clock of $O$ runs slow.

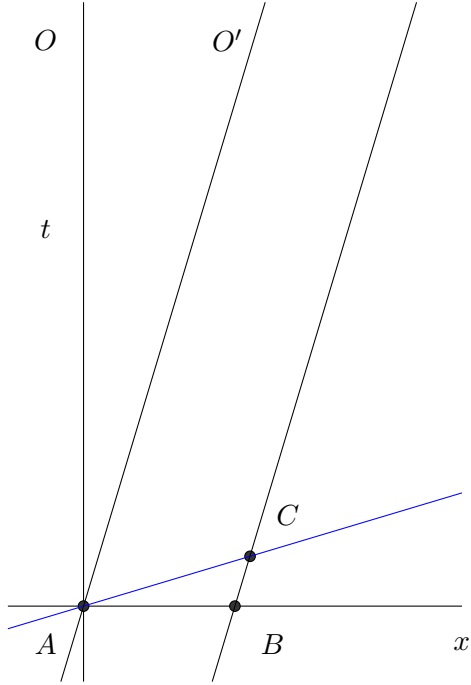
\begin{figure}
\centering
\begin{tikzpicture}
\draw (-1,0) -- (5,0);
\draw (0,-1) -- (0,8);
\node at (5,-0.5) {$x$};
\node at (-0.5,5) {$t$};
\node at (-0.5,7.5) {$O$};
\node at (1.9,7.5) {$O'$};
\node at (-0.5,-0.5) {$A$};
\node at (2.5,-0.5) {$B$};
\node at (2.7,1.2) {$C$};
\draw[fill=black!80] (0,0) circle (0.07);
\draw[fill=black!80] (2,0) circle (0.07);
\draw[fill=black!80] (2.2,0.66) circle (0.07);
\draw[blue] (-1,-0.3) -- (5.0,1.5);
\draw (-0.3,-1) -- (2.4,8);
\draw (1.7,-1) -- (4.4,8);
\end{tikzpicture}
\caption{An observer $O$ ($t$ axis) and a moving ruler}
\label{ruler}
\end{figure}

We now turn to the phenomenon of length contraction.  That is we consider a ruler whose length in its own rest frame is $L_0$ and ask what length $L$ it has when moving at velocity $v$. We begin by drawing a spacetime diagram of the moving ruler and the stationary observer $O$, see figure (\ref{ruler}).  
Part of the diagram is the same as the previous one in which we have the observer $O$ whose worldline is the $t$ axis and the observer $O'$ whose worldline is the line $x=vt$.  However, in this case that worldline is also the back of the ruler, and in additon we have the worldline of the front of the ruler $x=L+vt$.  Now let's consider how $O$ would answer the question ``what is the length of the ruler?''  The history of the ruler consists of a whole worldsheet consisting of all events between the back of the ruler and the front of the ruler.  Thus to pose the question of the length of the ruler, we first need to pick two events, in this case $A$ and $B$ representing the back and front of the ruler {\emph {at the same time}}.  The length of the ruler would then be the spatial distance between those two events as given by the square root of the spacetime interval.  Since the coordinates of $A$ are $(0,0)$ and those of $B$ are $(0,L)$ we find that the length according to $O$ is ${\sqrt {\Delta {x^2} - \Delta {t^2}}} = 
{\sqrt {{L^2} - 0}} = L$.  Now let's use the same method to see what length $L_0$ is assigned to the ruler by the observer $O'$ who is at rest with respect to the ruler.  The first task is to pick two events, one on the back of the ruler and one on the front of the ruler, that are at the same time.  But in this case by ``at the same time'' we mean ``at the same time according to $O'$'' {\emph {not}} ``at the same time according to $O$.''  From the formula ${t'}=\gamma(t-vx)$ we can see that ${t'}=0$ at event $A$.  Therefore for $O'$ the events that are at the same time at $A$ are those with ${t'}=0$ and therefore those with $t=vx$.  Thus for $O'$ the event at the front of the ruler that is at the same time as $A$ is event $C$, not event $B$.  Thus we see immediately that the main reason that $O$ and $O'$ disagree about the length of the ruler is that they are not even talking about the same thing: $O$ is talking about the length of the line segment $AB$ and $O'$ is talking about the length of the line segment $AC$.  We now calculate $L_0$ which is the length of the segment $AC$.  We start by finding the coordinates $({t_C},{x_C})$ of event $C$.  Since this point is on the line $t=vx$ and also on the line $x=L+vt$ we find ${t_c}=Lv/(1-{v^2})$ and ${x_c}=L/(1-{v^2})$.  Since length is the square root of the spacetime interval, we find that the length of the segment $AC$ is
\be
{L_0}={\sqrt {\Delta {x^2} - \Delta {t^2}}} = {\frac L {1-{v^2}}} {\sqrt {1-{v^2}}} = \gamma L
\; \; \; ,
\ee    
or in other words that 
\be
L = {L_0}/\gamma = {L_0} {\sqrt {1-{v^2}}} \; \; \; ,
\ee
so ``fast rulers become short'' or at least are measured to be that way by an observer with respect to whom they are moving.

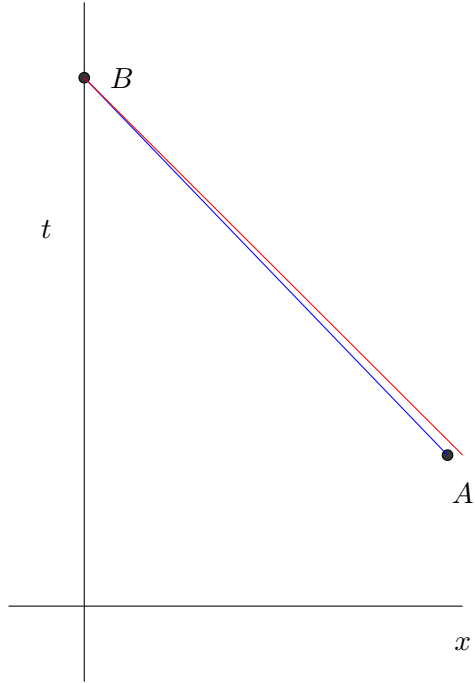
\begin{figure}
\centering
\begin{tikzpicture}
\draw (-1,0) -- (5,0);
\draw (0,-1) -- (0,8);
\node at (5,-0.5) {$x$};
\node at (-0.5,5) {$t$};
\node at (5,1.5) {$A$};
\node at (0.5,7) {$B$};
\draw[fill=black!80] (4.8,2) circle (0.07);
\draw[fill=black!80] (0,7) circle (0.07);
\draw[blue] (4.8,2) -- (0,7);
\draw[red] (5,2) -- (0,7);
\end{tikzpicture}
\caption{muon production and detection}
\label{muon}
\end{figure}

With all the emphasis on points of view of different observers, one might wonder whether there are any concrete, objective consequences of time dilation or length contraction.  So we now turn to one: the properties of a subatomic particle called the muon.  This particle can be thought of as a heavier version of the electron, but it is unstable and undergoes radioactive decay to an electron, a neutrino, and an antineutrino.  In its own rest frame the muon has a lifetime $\tau_\mu$ of about $2.2$ microseconds.  With that lifetime, Newtonian physics would predict that even at the speed of light a muon would only get to a distance of 
\be
c {\tau_\mu} = (3.0 \times {{10}^8} \, {\rm m}/{\rm s} )\, (2.2 \times {{10}^6} \, {\rm s}) = 660 \, {\rm m} \; \; \; .
\ee
And yet muons produced in the upper atmosphere by cosmic rays can still travel to ground level (a distance of about 100 km) and be detected there.  How can this be?  The answer (or as we will see, an answer) is time dilation: to get to ground level the muon will have to travel about 150 times as far as expected and therefore must have a lifetime about 150 times as much as expected.
This greater lifetime is achieved by traveling with a speed that gives $\gamma =150$.  Note however that we could just as easily explain this phenomenon using length contraction and the rest frame of the muon as follows: the muon is produced at one end of the atmosphere and lasts for $2.2$ microseconds.  In that time an atmosphere of thickness 660 m can rush by it.  Thus the Earth's atmosphere, with a rest thickness of 100 km must be contracted by a factor of 150 and therefore must be traveling with a speed that gives $\gamma =150$.  Alternatively, we can produce an explanation with a spacetime diagram and proper time, without ever talking about either time dilation or length contraction.  We do this as follows: consider the spacetime diagram of figure (\ref{muon})
Here the blue line represents the world line of the muon, with event $A$ being where the muon is produced in the upper atmosphere and event $B$ being where the muon is detected on the ground.  For comparison, the red line is the worldline of a light ray.  Despite the fact that these two lines are very close, we will see that the difference between the two lines is actually exagerated in the figure.  We demand that the proper time elapsed in the muon's world line equal the muon lifetime.  Since light rays have zero proper time, it is clear from the diagram that we can make the proper time as small as we like just by choosing the blue line sufficiently close to the red line.  With the muon traveling at speed $v$ we have $\Delta x /\Delta t = v$ and therefore
\be
{\tau_\mu} = {\sqrt {\Delta {t^2} - \Delta {x^2}}} = \Delta x {\sqrt {{\frac 1 {v^2} } - 1}}
= {\frac {\Delta x} {\gamma v}} \; \; \; .
\ee
Since $\Delta x/{\tau_\mu} \approx 150$ we have that $\gamma v \approx 150$ and therefore that
$\gamma \approx 150$.

\section{Paradoxes}
\label{paradox}
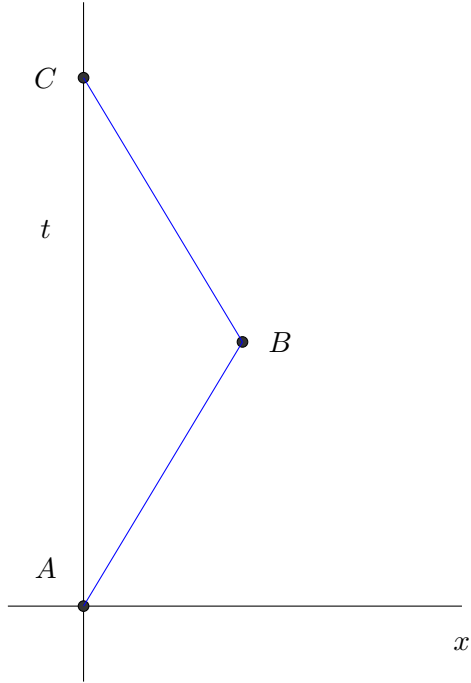
\begin{figure}
\centering
\begin{tikzpicture}
\draw (-1,0) -- (5,0);
\draw (0,-1) -- (0,8);
\node at (5,-0.5) {$x$};
\node at (-0.5,5) {$t$};
\node at (-0.5,0.5) {$A$};
\node at (2.6,3.5) {$B$};
\node at (-0.5,7.0) {$C$};
\draw[fill=black!80] (0,0) circle (0.07);
\draw[fill=black!80] (2.1,3.5) circle (0.07);
\draw[fill=black!80] (0,7) circle (0.07);
\draw[blue] (0,0) -- (2.1,3.5);
\draw[blue] (2.1,3.5) -- (0,7);
\end{tikzpicture}
\caption{Twins take different paths through spacetime and find that they are no longer the same age}
\label{twin}
\end{figure}

We now consider some paradoxes of special relativity.  However, we will take the point of view that there is very little that is paradoxical about them, and that viewed the right way ({\it i.e.} with spacetime diagrams and invariant quantities) the answers become rather obvious.  We begin with the twin paradox.  Recall from figure (\ref{catch}) that each observer thinks that the other's clock is running slow.  Furthermore, after the original synchronization of the two clocks at event $(0,0)$ the two observers never meet again, thus preventing any direct comparison of their two clocks to see whose clock is really the slow one.  But what if they did meet again?  Specifically, suppose that $O'$ instead of always receding at speed $v$ were to stop at some point and turn around, returning at speed $v$.  What then would a direct comparison of the two clocks show?  Put this way, this is a straightforward question that we can resolve in a straightforward way: draw a spacetime diagram and calculate the answer.  The diagram is given in figure (\ref{twin}).  For concreteness, suppose our observers are twins who at event $A$ are celebrating their 25th birthday.  One twin stays at home, while the other goes at 3/5 the speed of light to a star that is 6 light years away (reaching it at event $B$) and then comes back at the same speed.  How old is each twin when they meet (at event $C$)?  For the purpose of this diagram we will use units where time is measured in years and distance in light years.  We begin by finding the coordinates (in the $tx$ coordinate system) of each event, using $(0,0)$ as the coordinates of event $A$.  For event $B$ the coordinates are $(10,6)$, while for event $C$ they are $(20,0)$.  This immediately tells us that for the twin who remains home 20 years have elapsed, and when they meet this twin is 45 years old.  What about for the twin who goes to the star?  The proper time elapsed between events $A$ and $B$ is 
$\Delta \tau = {\sqrt {\Delta {t^2}-\Delta {x^2}}} = {\sqrt {{{10}^2} - {6^2}}} = 8$, while similarly the proper time elapsed between events $B$ and $C$ is 
$\Delta \tau = {\sqrt {\Delta {t^2}-\Delta {x^2}}} = {\sqrt {{{10}^2} - {{(-6)}^2}}} = 8$
Thus when the twins meet again the one who went to the star has an age of $25 +16=41$, definitely younger than his 45 year old twin.    
OK, so we asked a simple question and answered it with a simple picture and a simple calculation.  So what's so paradoxical?  Well for one thing the notion that twins can be two different ages.  I suggest approaching that part of the paradox with the slogan ``clocks are like odometers.''  Here's what I mean by that.  Suppose that figure (\ref{twin}) were a map in space rather than spacetime.  Let's say that $A$ and $C$ are two cities, say Detroit and Chicago.  And suppose you drive the direct route from Detroit to Chicago ($A$ to $C$), while I drive an indirect route ($A$ to $B$ to $C$).  Each of our car odometers will tell us how many miles we drove, and not suprisingly my odometer reading will be higher than yours.  That's because odometers don't measure the distance between two cities, but rather the length of the particular route that we drive.  Similarly, clocks measure proper time, which depends on the route we take through spacetime.  We're all familiar with the notion that the shortest distance between two points in space is a straight line.  Analogously, it turns out that in spacetime the longest proper time between two events is a straight line, {\it i.e.} the history of an inertial observer.  Thus the twin who goes away and comes back will always be younger than the twin who stays home.  The other thing that seems paradoxical about the twin paradox is that it seems to contradict the hard won notion of the previous section that there is no fact of the matter about whose clock runs slow: we have done a direct comparison between the two clocks and one of them is definitely slower.  The answer to this issue is that ``no fact of the matter about whose clock is slow'' is a statement about inertial observers.  The twin who stays home is an inertial observer, but the twin who goes out and back is not.  One can see this from the fact that this twin must accelerate in order to come back.  Alternatively, one can note that there is an inertial frame for the twin going out and there is an inertial frame for the twin coming back, but these are two {\emph {different}} inertial frames.  There is nothing that can be called the frame of the twin who goes out and comes back and therefore no contradiction with special relativity in the fact that the time on his clock is definitely different from that of his twin.

Just as the twin paradox comes from trying to answer the question of ``whose clock is really slow?'' so the pole in the barn paradox comes from trying to answer the question of ``whose ruler is really short?''  Suppose a pole vaulter runs into a barn with his pole which is longer than the barn is deep.  Since his pole is longer than the barn, it won't fit in the barn.  But suppose that we have him run so fast that length contraction makes his pole shorter than the barn.  Then it should fit into the barn, shouldn't it?  But wait, let's consider that same situation in the rest frame of the pole.  Then it is the barn that shrinks while the pole doesn't, so the pole should definitely not fit in the barn.    
\begin{figure}
\centering
\begin{tikzpicture}
\draw (-1,0) -- (5,0);
\draw (0,-1) -- (0,8);
\draw (4.5,-1) -- (4.5,8);
\node at (5,-0.5) {$x$};
\node at (-0.5,5) {$t$};
\node at (-0.5,0.5) {$A$};
\node at (5,0.83) {$B$};
\draw[fill=black!80] (0,0) circle (0.07);
\draw[fill=black!80] (4.5,0.83) circle (0.07);
\draw[blue] (0,0) -- (4.8,8);
\draw[blue] (4,0) -- (4.9,1.5);
\draw[red] (-0.5,-0.3)  -- (5,3);
\end{tikzpicture}
\caption{Does the pole fit in the barn?}
\label{pole}
\end{figure}
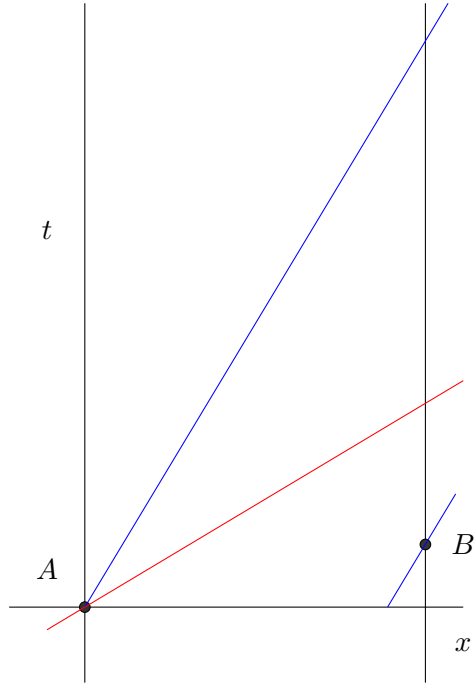
For concreteness and ease of calculation, we will use particular numbers similar to the ones in the previous example.  Suppose that the rest length of the pole is 50 feet and that of the barn is 45 feet, and that the pole vaulter runs at a speed of 3/5.  As shown in the previous calculation $v=3/5$ corresponds to $\gamma = 5/4$.  So from the barn's point of view the length of the pole is $50/\gamma=40$ feet, so the pole will fit into the 45 foot barn.  However, from the point of view of the pole, the barn is $45/\gamma=36$ feet, so the 50 foot pole will definitely not fit into the 36 foot barn. So who is right?  Does the pole fit in the barn or doesn't it? 
To resolve this question we will draw a spacetime diagram, figure (\ref{pole}).
Here the $t$ axis represents the world line of the front of the barn, while the other vertical line represents the world line of the back of the barn.  The blue line through the origin represents the world line of the back of the pole, while the other blue line represents the world line of the front of the pole.  Event $A$ is the event of the back of the pole entering the barn through the front, while event $B$ is the event of the front of the pole leaving the barn through the back.  Thus the somewhat vague question ``does the pole fit in the barn?'' can be rendered more precisely as ``is there some time when the whole pole is inside the barn?'' which in turn can be posed completely precisely as ``does event $B$ occur after event $A$?'' At first sight the answer seems to be yes, because event $A$ occurs at $t=0$ and event $B$ occurs at $t>0$. But on closer inspection this only means that the pole fits in the barn from the perspective of the barn, which is what we already expected.  Instead we could look at the notion of simultaneity associated with the pole.  Here the set of events simultaneous with $A$ is the $x'$ axis given by $0={t'}=\gamma(t-vx)$ and therefore $t=vx$, the red line in the figure.  Event $B$ is below the red line and therefore in the frame of the pole event $B$ happens before event $A$.  More generally, we can simply calculate the spacetime interval between $A$ and $B$.  Pairs of events for which $\Delta {s^2} > 0$ are called spacelike separated, and there is no fact of the matter as to whether one event is before the other.  Event $B$ is on the line $x=40+(3/5)t$ and also on the line $x=45$.  Therefore its $(t,x)$ coordinates are $(25/3,45)$.
Thus the spacetime interval is $\Delta {s^2} = {{45}^2}-{{25/3}^2} \approx 1955 > 0$.  Thus there is no fact of the matter as to whether $A$ is before $B$ and therefore there is no fact of the matter about whether the pole fits inside the barn.  We then see that this paradox can be resolved using only a spacetime diagram and the spacetime interval.

For our final paradox we will consider velocity addition.  Recall that special relativity forbids motion faster than the speed of light.  But suppose that you are standing still and I am moving at 3/4 of the speed of light and I throw a baseball in the direction of my motion at 3/4 of the speed of light relative to me.  Then it sounds like the baseball should move at (3/4+3/4)=3/2 of the speed of light relative to you. The usual way to resolve this paradox is to use the boost that relates my frame to yours on the motion of the baseball and thus find an expression for the speed of the baseball in your frame.  Instead we will resolve it by using the fact that boosts are like rotations.  Recall that if I rotate by an angle $\alpha _1$ and then rotate by an angle $\alpha_ 2$, that is the same as rotating by an angle ${\alpha_1}+{\alpha_2}$.  Similarly if my velocity relative to you is $v_1$ and the velocity of the baseball relative to me is $v_2$ then the velocity of the baseball relative to you is $\tanh ({\psi_1}+{\psi_2})$ where ${v_1}=\tanh {\psi_1}$ and ${v_2}=\tanh {\psi_2}$. Note that this already tells us that the speed of the baseball is less than the speed of light because $\tanh (\psi) < 1$ for any $\psi$.  Nonetheless, we would also like to have a formula for that velocity in terms of $v_1$ and $v_2$.  To obtain that formula we first note that there are several trigonometric identities involving the sum of angles.  One of these identities is 
\be
\tan ({\alpha _1}+{\alpha_2})={\frac {\tan {\alpha _1}+\tan {\alpha _2}} {1-\tan {\alpha _1}\tan {\alpha _2}}} \; \; \; .
\ee
The corresponding identity for hyperbolic tangent is
\be
\tanh ({\psi _1}+{\psi_2})={\frac {\tanh {\psi _1}+\tanh {\psi _2}} {1+\tanh {\psi _1}\tanh {\psi _2}}} \; \; \; .
\ee
Therefore the velocity of the baseball relative to you is
\be
{\frac {{v_1}+{v_2}} {1+{v_1}{v_2}}} \; \; \; .
\ee
In particular, if $v_1$ and $v_2$ are each 3/4, then the relative velocity is
$(3/2)/(1+9/16)=24/25$ which is less than 1.

\section{Towards General Relativity}
\label{GR}

The bravest of retired engineers don't stop at reformulating special relativity: they also want to reformulate general relativity.  In this final section of the paper we will first consider to what extent the insights of the previous sections for how to understand special relativity can be carried forward to also enhance understanding of general relativity.  Much of the advice for what {\emph {not}} to do can simply be carried forward with very little change: (1) don't naively use Babylonian mathematics.  It has its pitfalls.  (2) don't trust your physical intuition, especially for a theory whose natural regime (high velocity and strong gravity) is far from anything you have encountered in the lab.  (3) don't make up your own axioms under the mistaken impression that ``that's how the game is played.'' It isn't.  Furthermore if you make up axioms without the ability to calculate their consequences and compare to experiment then you have no way of finding out that your original idea is wrong and you should try something else: {\emph {not being able to find out when you're wrong is an almost guaranteed recipe for failure}}. (4) Don't try to put everything in Newtonian terms.  Instead learn general relativity on its own terms.

In addition to the ``don'ts'' there are some positive lessons that carry forward: (1) doing detailed calculations using boosts to go from one frame to another is not the heart of special relativity.  That's a good thing because there are no global frames in general relativity and therefore no way to carry forward that activity to general relativity. Instead, special relativity is about the behavior of clocks and light as determined by the flat spacetime metric $\eta _{\alpha \beta}$.  Correspondingly general relativity is about the behavior of clocks and light as determined by the curved spacetime metric $g_{\alpha \beta}$.  (2) A coordinate system is a needed setup for doing calculations, but the coordinates themselves don't have a direct physical meaning.  One finds out their meaning by working out the behavior of clocks and light.  (3) The twin who stays at home has the longest proper time.  Inertial observers are those who move subject to no force, and also those who maximize the proper time between any two points on their world line.  Thus one could use proper time as the action in the principle of least action to derive the equation of motion for inertial observers.  This fact, which at first seems like a small sideline in special relativity, acquires central importance in general relativity as follows: there are no inertial observers in general relativity, but the closest thing to an inertial observer is one who is in free fall, that is one who is acted on by no forces except gravity.  Finding these free fall trajectories is a big part of general relativity.

Perhaps the largest obstacle to retired engineers learning general relativity is the mathematics involved, particularly differential geometry.  This part of mathematics is no part of the education of an engineer, nor of the education of the majority of physicists.  The retired engineer has gotten along all his professional life without abstruse mathematics and doesn't intend to change.  Because of this retired engineers interested in general relativity are drawn to the principle of equivalence, which says that locally being at rest in a uniform gravitational field is equivalent to being under constant acceleration in special relativity.  In principle this sounds great because then one can leverage special relativity calculations to say something about gravity.  Retired engineers develop the misimpression that somehow special relativity can't properly deal with acceleration and that in any case understanding gravity is about figuring out what the frame of an accelerated observer is.  It is important for retired engineers to understand just how severly limitted the principle of equivalence is.  It is helpful to recall that Einstein came up with the principle of equivalence in 1907, but it took him until the end of 1915 to come up with general relativity.  Along the way he produced several different versions, some of which made wrong predictions for the bending of light, or the perihelion shift of Mercury, or both.  The final version of the theory very much depends on differential geometry for its underlying structure and for the calculation of its physical effects.  Thus, one of the main things that retired engineers need to understand about general relativity is that there's more to life than the principle of equivalence.

What hope then is there for retired engineers who want to understand general relativity but don't want to learn differential geometry?  One promising approach is Jim Hartle's ``physics first'' method.\cite{hartle}  To explain how this method works, it is helpful to recall John Wheeler's pithy summary of general relativity as, ``matter tells spacetime how to curve, and spacetime tells matter how to move.''  In this terminology Hartle's pedagogical insight is that the mathematics behind ``matter tells spacetime how to curve'' can be very abstract and complicated, whereas the mathematics behind ``spacetime tells matter how to move'' is comparatively simple.  The ``physics first'' approach is then to cover the ``spacetime tells matter how to move'' topics first, introducing almost no new mathematics to do so, and only later to cover the ``matter tells spacetime how to curve'' part, introducing the complicated mathematics only when needed. 

In particular, without introducing any new mathematics one can think of the spacetime metric $g_{\alpha \beta}$ as a matrix.  The inverse metric $g^{\alpha \beta}$ is then just the inverse of that matrix.  ``Objects in free fall maximize proper time,'' is then just another case in which the principle of least action is used to obtain equations of motion.  The equation of motion that one obtains in this way is
\be
{\frac {d{u^\gamma}} {dt}} + {\Gamma ^\gamma _{\alpha \beta }}{u^\alpha}{u^\beta} = 0 \; \; \; .
\label{geodesic}
\ee
Here the quantities ${\Gamma ^\gamma _{\alpha \beta }}$ are called the Christoffel symbols and are given by the expression
\be
{\Gamma ^\gamma _{\alpha \beta }} = {\textstyle {\frac 1 2}} {g^{\gamma \delta}} \left ( 
{\partial _\alpha}{g_{\beta \delta}}+{\partial _\beta}{g_{\alpha \delta}} -
{\partial _\delta}{g_{\alpha \beta}} \right ) \; \; \; .
\label{Christoffel}
\ee
Equation (\ref{Christoffel}) is sufficiently messy that one of my professors liked to joke that the Christoffel symbols should really be called the ``Christ-Awful symbols.''  Luckily there are some cases of great physical interest in which one can avoid the Christoffel symbols altogether.  This is because some metrics have symmetry, and each symmetry leads to a conserved quantity.  In general relativity this works as follows: suppose that the metric components are independent of one of the coordinates, which we will call $x$.  Let the vector field $\xi ^\alpha$ have ${\xi ^x}=1$ with all other components zero.  Then ${\xi ^\alpha}{u_\alpha}={u_x}$ is a constant.

We will end with an example of this method.  The infinitesimal spacetime interval takes the form \be
d {s^2} = {g_{\alpha \beta}}d{x^\alpha}d{x^\beta} \; \; \; .
\label{lineelement}
\ee
For the Schwarzschild metric this infinitesimal spacetime interval is
\be
d {s^2} = - \left ( 1 - {\frac {2M} r} \right ) d {t^2} + {{\left ( 1 - {\frac {2M} r} \right ) }^{-1}} d {r^2} + {r^2} (d {\theta ^2} + {\sin^2}\theta d {\phi^2}) \; \; \; .
\label{schwarzschild}
\ee   
This metric describes the gravitational field outside any spherical object, and also describes all non-rotating black holes.  Here the constant $M$ is the mass of the object or black hole and we are using units in which not only $c$ but also Newton's gravitational constant $G$ are set equal to 1.
Comparing eqn. (\ref{lineelement}) to eqn. (\ref{schwarzschild}) we find that the nonzero components of the metric are 
\be
{g_{tt}} = - \left ( 1 - {\frac {2M} r} \right ) , \; \; \; {g_{rr}} =  {{\left ( 1 - {\frac {2M} r} \right ) }^{-1}} 
, \; \; \; {g_{\theta \theta}} = {r^2} , \; \; \; {g_{\phi \phi}} = {r^2} {\sin^2}\theta \; \; \; .
\label{schwarzschild2}
\ee
The metric components are independent of $t$, so there is a constant $E$ such that ${u_t}=-E$.  Since ${u_t}={g_{tt}}{u^t}$ we also have 
\be
{u^t} = E {{\left ( 1 - {\frac {2M} r} \right ) }^{-1}} \; \; \; .
\ee
The metric components are independent of $\phi$, so there is a constant $L$ such that 
${u_\phi}=L$. Since ${u_\phi}={g_{\phi \phi}}{u^\phi}$ we have 
${u^\phi}=L/({r^2}{\sin^2}\theta)$.  We will choose the plane of the orbit to be $\theta=\pi/2$ so that ${u^\theta}=0$ and $\sin \theta = 1$.  We then find
\be
-1={u^\alpha}{u_\alpha} = {u^t}{u_t}+{g_{rr}}{{({u^r})}^2} + {u^\phi}{u_\phi}
= {{\left ( 1 - {\frac {2M} r} \right ) }^{-1}} \left [ - {E^2} + {{({u^r})}^2} \right ] 
+ {\frac {L^2} {r^2}} \; \; \; .
\label{geodesic2}
\ee
However, ${u^r} = dr/d\tau = {\dot r}$ where we are using an overdot to denote derivative with respect to $\tau$.  We then find that eqn. (\ref{geodesic2}) takes the form
\be
{{\dot r}^2} + {\tilde V}(r) = {E^2} \; \; \; ,
\label{geodesic3}
\ee 
where ${\tilde V}(r)$ is given by
\be
{\tilde V}(r) = \left ( 1 - {\frac {2M} r} \right ) \left ( 1 + {\frac {L^2} {r^2}} \right )
\; \; \; .
\ee
Note that eqn. (\ref{geodesic3}) is the equation for one dimensional motion of a particle of mass 2 in a potential ${\tilde V}(r)$ with energy $E^2$.  Thus orbital motion in general relativity is reduced in this way to a simple mechanics problem.  We therefore see that there are aspects of general relativity that are easily accessible even to retired engineers.

\section*{Acknowledgements}
To those people reading this manuscript who also know Bob Wald and Bob Geroch, their influence will be obvious in every line.  Nonetheless, neither of the Bobs should be held responsible for any of the views expressed here.

\end{document}